\documentclass[aps,prb,twocolumn,superscriptaddress,amsmath,amssymb]{revtex4-2}
\usepackage{newtxtext,newtxmath}
\usepackage{graphicx}
\usepackage{bm}
\usepackage{siunitx}
\usepackage{tabularx}
\usepackage[colorlinks=true,citecolor=blue,breaklinks=true,linkcolor=blue,
linktocpage=true,urlcolor=blue,pagebackref=false]{hyperref}
\usepackage{appendix}

\DeclareGraphicsExtensions{.pdf,.png}
\newcommand{\up}{{\uparrow}}
\newcommand{\down}{{\downarrow}}

\newcommand{\asection}[1]{\section{\uppercase{#1}}}

\begin{document}
	\title{Direct measurement of spin-flip rates in single-electron tunneling}
	
	\author{Olfa Dani}
	\affiliation{Institut f\"ur Festk\"orperphysik, Leibniz Universit\"at Hannover, D-30167 Hanover, Germany}
	
	\author{Robert Hussein}
	\affiliation{Institut f\"ur Festk\"orpertheorie und -optik, Friedrich-Schiller-Universit\"at Jena, D-07743 Jena, Germany}
	
	\author{Johannes C. Bayer}
	\affiliation{Institut f\"ur Festk\"orperphysik, Leibniz Universit\"at Hannover, D-30167 Hanover, Germany}
	
	\author{Klaus Pierz}
	\affiliation{Physikalisch-Technische Bundesanstalt, Bundesallee 100, D-38116 Braunschweig, Germany}
	
	\author{Sigmund Kohler}
	\affiliation{Instituto de Ciencia de Materiales de Madrid, CSIC, E-28049 Madrid, Spain}
	
	\author{Rolf J. Haug}
	\affiliation{Institut f\"ur Festk\"orperphysik, Leibniz Universit\"at Hannover, D-30167 Hanover, Germany}
	
	\date{\today}

	\begin{abstract}
		
		Spin-flips are one of the limiting factors for spin-based information
		processing. We demonstrate a transport approach for determining the
		spin-flip rates of a self-assembled InAs double quantum dot occupied by a
		single electron. In such devices, different Land\'e factors lead to an
		inhomogeneous Zeeman splitting, so that the two spin channels can never be
		at resonance simultaneously, leading to a spin blockade at low
		temperatures. This blockade is analyzed in terms of spin flips for
		different temperatures and magnetic fields. Our results are in good
		agreement with a quantum master equation that combines the dot-lead
		couplings with ohmic dissipation stemming from spin-flip cotunneling.
		
	\end{abstract}
	
	\maketitle
	
	\section{Introduction}
	The concept of quantum computing based on spins in coupled quantum dots was
	proposed 25 years ago \cite{LossPRA98} and appears to be within reach of
	applications nowadays \cite{WeinsteinNL23, GilbertNN23, BurkardRMP23}. Initialization and
	control of the spin is critically dependent on the coherence of the spins. 
	A number of theory works showed a sensitive dependence of the spin coherence
	on the hyperfine coupling \cite{SchliemannPRB02} or cotunneling
	\cite{QassemiPRL09, CoishPRB11}, as well as on the
	influence of Markovian \cite{CoishPRB08} or non-Markovian \cite{CoishPRB04}
	noise.  The spin-phonon coupling mediated by spin-orbit interaction is
	often quite small and the dominating contribution to spin decoherence is
	generally assumed to happen via hyperfine interaction. Spin relaxation
	rates were addressed experimentally for single quantum dots
	\cite{HansonPRL03, ElzermanNL04, AmashaPRL08, CamenzindNC18, KurzmannPRL19,
	BanszerusNC22} and also for coupled quantum dots \cite{OnoS02, JohnsonNL05,
	KoppensPRL08, BluhmNP11, SrinivasaPRL13, MaisiPRL16}.
	There the decay rate as a function of the Zeeman splitting hints
	on the dominating dissipation process.
	In coupled quantum dots, mostly the spin blockade mechanism
	\cite{OnoS02} based on the energetic difference between singlet and triplet
	two-electron states was studied in such measurements. It was shown that
	these two-electron states relax the spin predominantly via hyperfine
	interaction and spin-orbit interaction in III-V semiconductors
	\cite{NadjPergeNL10, WangNL18, SchroerPRL11}, whereas in  silicon and
	germanium dots also cotunneling \cite{LaiSR11, YamahataPRB12, LiNL15,
	BraunsPRB16, ZhangNL21} can be of importance. For transport phenomena that
	depend on the dynamics of a single spin in a double quantum dot only very
	few studies exist \cite{HuangPRL10, MaisiPRL16, DaniCP22}.
	
	Here, we demonstrate how to directly extract the spin-flip rates for
	single spins in coupled quantum dots from the measured single-electron
	tunneling current. To this end, we use a double quantum dot with a Zeeman
	level structure similar to the one of Ref.~\cite{HuangPRL10}, but with an
	inter-dot tunneling much smaller than the typical detuning. Then, in
	contrast to this former work, the resulting current blockade is resolved by
	spin flips.  For their theoretical description, we consider spin-flip
	cotunneling \cite{QassemiPRL09, CoishPRB11}, i.e., dissipative spin transitions
	accompanied by excitations at the Fermi surface of a lead.  As a
	consequence, the spin experiences ohmic dissipation \cite{LeggettRMP87,
	HanggiRMP90} which fits the experimental data rather well, i.e., we
	show that spin-flip cotunneling is the dominant mechanism for spin
	relaxation in our double quantum dot occupied by a single electron.
	
	\begin{figure}[b]
		\centerline{%
		\includegraphics[width=.95\columnwidth]{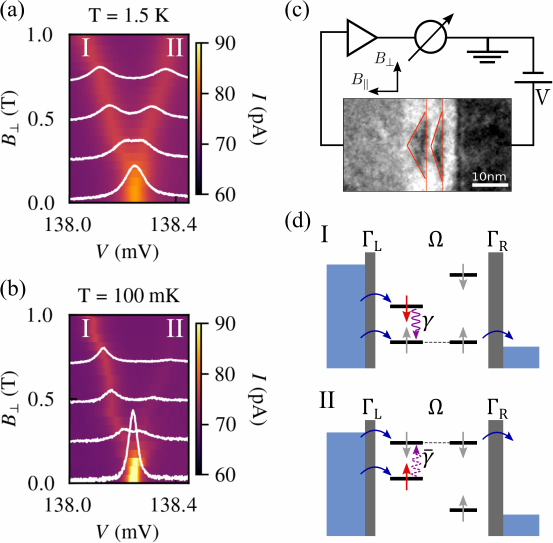}}
		\caption{
		(a) Current-voltage characteristic of InAs double quantum dots as function
		of the magnetic field applied perpendicular to the tunneling current up to
		$B = \SI{1}{T}$ at $T = \SI{1.5}{K}$, showing a current peak at $V \approx
		\SI{138.25}{mV}$ for $B=0$. The white lines depict cuts along the color
		graphs, showing the current peaks for a given magnetic field.
		(b) Same for $T = \SI{100}{mK}$.
		(c) TEM image of an InAs double quantum dot shown in a schematic picture of
		the measurement setup and of the different magnetic field directions used.
		(d) Sketch of the transport channels for the different situations
		of resonance I and II. Due to the inhomogeneous Zeeman splittings always
		one channel is off-resonant for the two different resonance conditions. An
		electron may be trapped in that channel and block transport until a spin flip
		indicated by the purple arrow brings the electron to the resonant channel.}
		\label{fig:1}
	\end{figure}
	
	\begin{figure}
		\centerline{%
		\includegraphics[width=.95\columnwidth]{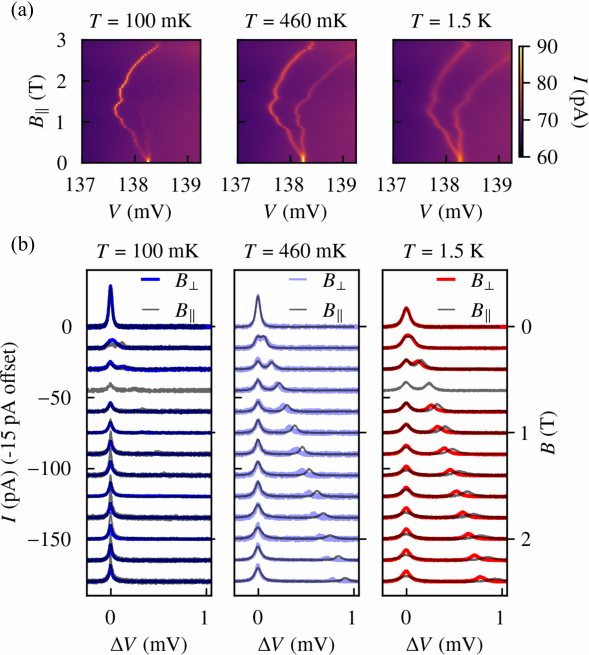}}
		\caption{(a) Magnetic field dependence of the tunneling current as function
		of voltage for magnetic field applied parallel to the tunneling current for
		three different temperatures $T = \SI{100}{mK}$, $\SI{460}{mK}$, and
		$\SI{1.5}{K}$.
		(b) Tunneling current as function of applied voltage normalized to
		the position of resonance I for the two different magnetic field
		directions. Traces offset by \SI{-15}{pA} are shown for magnetic fields
		ranging from 0 up to \SI{2.4}{T} in step sizes of \SI{0.2}{T}. 
		}
		\label{fig:2}
	\end{figure}
	
	\section{Experiment}
	For our investigation, we use vertically coupled self-assembled InAs
	quantum dots. The quantum dots are embedded in a GaAs-AlAs heterostructure.
	Due to the
	lattice-mismatch of InAs and AlAs pyramid-shaped quantum dots are formed
	and vertically aligned \cite{XiePRL95}. The dot in the second layer is
	slightly larger than the quantum dot in the first one \cite{EiseleAPL99}
	due to the change in strain field. The diameter of such quantum dots
	typically is 10 to \SI{20}{nm} and their height is between 2 and
	\SI{4}{nm}. In these small dots the strong confinement leads to the Land\'e
	g-factor deviating strongly from the bulk value, approaching $g=2$ of the
	free electron with decreasing dot sizes. The effective thickness of the
	middle and top barriers is reduced by the quantum dots partially
	penetrating the AlAs layers, which results in asymmetric dot-lead tunneling
	rates. A TEM image of such an InAs double quantum dot is depicted in
	Fig.~\ref{fig:1}(c). The device is similar to the ones used in Refs.\
	\cite{BartholdPRL06, KiesslichPRL07, DaniCP22}, i.e. for zero bias the
	quantum dot levels are above the Fermi energy and tunneling is not possible
	whereas for finite bias voltage the quantum dot levels are shifted by the
	electric field and brought into resonance and into the transport window.
	 
	The color graphs in Figs.~\ref{fig:1}(a,b) show the measured current $I$
	through the InAs double quantum dot as a function of bias voltage $V$ and
	magnetic field up to $B = \SI{1}{T}$ at $T = \SI{1.5}{K}$ and $T =
	\SI{100}{mK}$, respectively. The magnetic field was applied perpendicular to
	the current (parallel to the layer structure).  The graphs for $B = 0$ show a current
	peak at $V \approx \SI{138.25}{mV}\equiv V_0$, owing to the resonant tunneling of
	single electrons through the InAs double quantum dot. The resonance originates from
	tunnel cycles with the occupation $(0, 0)\to(1, 0)\to(0, 1)$, where
	only a single electron is present in the double quantum dot.  At $T = \SI{1.5}{K}$, the
	single resonance splits into two peaks (a left peak: peak I and a right
	peak: peak II) for increasing magnetic field.  At $T = \SI{100}{mK}$
	[Fig.~\ref{fig:1}(b)], the resonant tunneling peak amplitude at $B=0$ is
	higher \cite{DaniCP22} and decreases drastically
	already with small magnetic fields $\sim \SI{0.25}{T}$ applied.  The
	amplitude of peak I stays at a low but finite level with increasing
	magnetic field, meanwhile the right peak fades away totally.
	
	This observation can be explained by the different effect of spin
	relaxation as sketched in Fig.~\ref{fig:1}(d). 
	Due to the different sizes of the two dots, their g-factors $g_L$ and $g_R$
	are different and therefore the Zeeman splitting becomes inhomogeneous~\cite{BjorkPRB05,HapkeWurstPE02}. For
	$V < V_0$ (peak I) the spin-down Zeeman levels of both quantum dots are in
	resonance, such that we expect to find a current peak.  An electron with
	spin-up, however, entering the left dot will get stuck, because the right
	dot is off-resonance, such that the resonant channel becomes blocked. This
	blockade can be resolved by a spin flip with rate $\gamma$ for all
	temperatures because it does not require a spin excitation. For $V > V_0$
	(peak II) the spin-up Zeeman levels are in resonance and can generate a
	tunneling current. If a spin-down electron enters the left quantum dot, it
	blocks the single-electron tunneling current. This blockade can be lifted
	by a spin flip to the spin-up level with a rate $\bar\gamma$.
	However, its lifting requires spin excitation and, thus, is observed only
	at sufficiently high temperatures such that $kT$ exceeds the Zeeman
	splitting.  Therefore, we observe peak II at \SI{1.5}{K} whereas at
	\SI{100}{mK} it cannot be observed for magnetic fields larger than
	\SI{0.5}{T}.
	
	At this stage, it is worth emphasizing the difference to
	Ref.~\cite{HuangPRL10} where a similar level structure has been studied.
	There, the inter-dot tunneling $\Omega$ is two orders of magnitude larger
	and comes close to the Zeeman splittings and exceeds the dot-lead
	couplings (in our setup by contrast, $\Omega\ll\Gamma_{L,R}$).
	Then the spin-up as well as the
	spin-down states hybridize and form delocalized orbitals.  When all levels
	are within the voltage window, this causes a current peak at $V=V_0$ even
	in the absence of spin flips.  By contrast, when $\Omega$ is much smaller
	than the difference of the Zeeman splittings, we observe a
	blockade which is resolved by spin flips.  Hence the current
	peaks provide information on the spin-flip rates.
	
	The color graphs in Fig.~\ref{fig:2}(a) show the current
	as a function of the bias voltage and the magnetic field for various
	temperatures, where a magnetic field $B_{\parallel}$ was applied parallel
	to the current (perpendicular to the layer structure). The observed
	oscillations and shifts with magnetic field originate from the Landau-level
	structure in the emitter.  Nevertheless, the peak-to-peak distances of the
	double peak increase more or less linearly with the magnetic field as shown
	in Fig.~\ref{fig:2}(b), where the position of peak~II is shown relative to
	the position of peak~I. In the same way in Fig.~\ref{fig:2}(b) also the
	results for the other magnetic field direction are shown.
	In both cases, the observed difference in peak positions is given by
	\begin{equation}
		\Delta V = |g_L-g_R|\mu_BB / \eta ,
	\end{equation}
	with the gyromagnetic ratio $g_\ell$ of quantum dot $\ell=L,R$ and the
	leverage factor $\eta$. The individual Zeeman splittings  $g_\ell\mu_BB$ of
	the two dots do not appear, only the difference $\Delta g =g_L-g_R$ plays a
	role here. Taking into account the leverage factor of $\eta$ = 0.15 we
	obtain for $B_{\perp}$ a difference in the Land\'e factors $\Delta g=0.85$,
	whereas for $B_{\parallel}$ the value is $\Delta g = 1$.  Anisotropy of the
	g-factor is well known for single InAs quantum dots
	\cite{MeyerPSSB01,HapkeWurstPE02,SchwanAPL11,BelykhPRB16}.
	Thus, the observed difference in $\Delta g$ for the different magnetic
	field directions can be attributed to the anisotropy of the g-factors of the
	individual quantum dots.
	
	\section{Double quantum dot--environment model}
	For a theoretical description of our observations, we start with the model used in
	Ref.~\cite{DaniCP22} and also consider spin-flip cotunneling \cite{QassemiPRL09, CoishPRB11}.
	In doing so, we model each quantum dot with a single level with onsite
	energy $\epsilon_\ell$ ($\ell=L,R$) and tunnel coupling $\Omega$.  In
	the operating regime of our experiment, Coulomb repulsion allows the
	occupation of the double quantum dot with only one electron. Dot $L$ is tunnel coupled
	also to an electron source from which electrons may enter at rate
	$\Gamma_L$.  Correspondingly, dot $R$ is coupled to a drain with
	rate $\Gamma_R$.  The broadening of the current peaks with increasing
	temperature can be explained by the coupling of the double quantum dot dipole moment to a
	bosonic heat bath \cite{LeggettRMP87, HanggiRMP90, Weiss2012} with ohmic
	spectral density and a dimensionless coupling strength $\alpha$.
	For details see Ref.~\cite{DaniCP22}.
	
	Spin flips may be treated at different levels ranging from a
	phenomenological Lindblad master equation to microscopic models for
	spin-orbit interaction and the hyperfine interaction with the magnetic
	moments in the substrate.  Such models may predict characteristic
	dependencies of the decay rates on the Zeeman splitting.  In turn,
	measurements of the spin-flip rates hint on the dominating
	mechanism.  In the present case, a model must be capable of
	explaining the significant temperature dependence and the 
	observed increase of the spin-flip
	rate with the Zeeman splitting.  Moreover, it must provide the
	asymmetry between spin decay and thermal excitation.
	
	These requirements can be fulfilled by spin-flip cotunneling
	\cite{QassemiPRL09, CoishPRB11} induced by the spin-conserving tunnel
	Hamiltonian $H_T = \sum_{k\sigma} t_k (c^\dagger_{k\sigma} d_\sigma +
	d_\sigma^\dagger c_{k\sigma})$, where
	the fermionic operators $c_{k\sigma}$ and $d_\sigma$ annihilate an electron
	with spin $\sigma$ in the left lead and the left dot, respectively (spin
	flips in the right dot do not play a role and will be ignored).
	In a $T$-matrix formulation, the impact of $H_T$ is given by an effective
	coupling that obeys the recursive
	relation $T = H_T + H_T G_0(E_i) T$, where $G_0(z) = (z-H_0)^{-1}$ is the
	Green function in the absence of tunneling and $E_i$ the energy of the
	initial state \cite{BruusFlensberg2004}.  
	
\begin{figure}
		\centerline{%
		\includegraphics[]{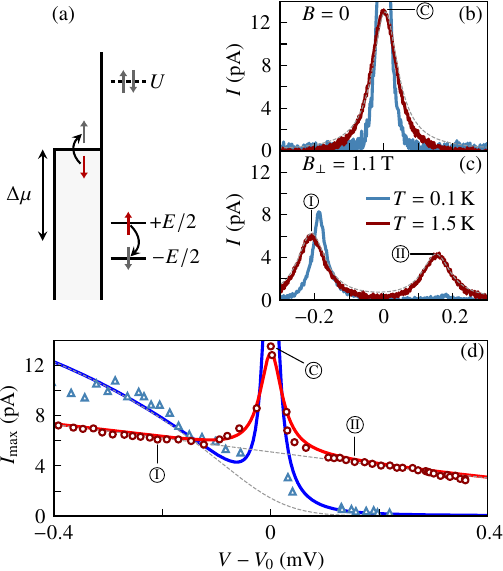}} %
		\caption{\label{fig:ExpTheo}
		(a) Spin dissipation due to cotunneling.
		(b,c) Measured tunneling  current as a
		function of the bias voltage in the absence ($B=0$) and presence
		($B_\perp=\SI{1.1}{\tesla}$) of a magnetic field, respectively, for
		temperatures $T=\SI{100}{mK}$ and $T=\SI{1.5}{K}$. The resonance peaks are
		centered around the peak position $V_0$ at $B=0$. The dashed line indicates
		the numerical  data for the higher temperature with
		$\Omega=\SI{0.85}{\micro\eV}$, $\Gamma_R=\SI{5}{\micro\eV}$,
		$\alpha=0.005$, $\alpha_\text{spin}=1.2\times10^{-4}$, and $g_L=0.7$.
		(d) Corresponding peak heights as a function of the bias voltage.
		Symbols mark the experimental values, while the solid lines 
		indicate numerical data. The dotted lines outline the analytical
		approximation in Eq.~\eqref{Ianalyt}.}
	\end{figure}

	Our focus lies on spin flips in a singly occupied quantum dot, where
	Coulomb repulsion and Pauli principle forbid resonant dot-lead tunneling,
	such that the leading contribution to $T$ is the second-order term $H_T G_0
	H_T$ which causes the process schematically shown in
	Fig.~\ref{fig:ExpTheo}(a).  A spin-up electron flips
	to the lower Zeeman level, while a lead electron at the Fermi
	surface is excited.  The opposite process requires the decay of a lead
	electron, which is possible only at sufficiently high temperature.  The
	virtually populated dot states of this second-order process are the empty
	dot and the singlet state.  In the Appendix, we show that
	the $T$-matrix can be approximated as $T = \hbar(\sigma^\dagger_\up\sigma_\down \zeta
	+ \sigma^\dagger_\down\sigma_\up \zeta^\dagger)$, where $\zeta$ describes
	quantum noise stemming from electron excitations under spin-flip at the
	Fermi surface of the lead.  For the numerical treatment,
	we employ a Bloch-Redfield master equation \cite{RedfieldIBMJRD57,
	Blum1996}.  The dissipative kernel stemming from $T$ is fully specified by the
	noise correlation $C(t) =
	\langle\zeta(0)\zeta^\dagger(t)\rangle$ which in frequency space reads
	$C(\omega) = \pi\alpha_\text{spin}\omega n_\text{th}(\hbar\omega)$ with the
	Bose function $n_\text{th}(E) = [\exp(E/kT)-1]^{-1}$.  While the order of
	magnitude of the dimensionless dissipation strength $\alpha_\text{spin}$
	can be estimated from the dot-lead coupling $\Gamma_L$, the onsite
	interaction, and the chemical potential, we will
	determine its precise value by fitting our experimental data.
	Details of the derivation and the relation to the ohmic spin-boson model
	are provided in the Appendix.
	
	Before focussing on spin flips, we consider the current peaks in the
	absence of the magnetic field shown in Fig.~\ref{fig:ExpTheo}(b).
	Proceeding as in Ref.~\cite{DaniCP22}, we determine the dot-lead rates
	$\Gamma_{L/R}$, the inter-dot tunneling $\Omega$, and the dimensionless
	dissipation strength $\alpha$ of the orbital degree of freedom to
	$\Omega=\SI{0.85}{\micro\eV}$, $\Gamma_R=\SI{5}{\micro\eV}$,
	$\alpha=0.005$. Due to the asymmetric coupling, $\Gamma_R \ll \Gamma_L$
	while the value of $\Gamma_L$ does not greatly influence the current. In
	comparison to the dots studied in Ref.~\cite{HuangPRL10} this is two orders
	of magnitude smaller inter-dot tunneling and even in comparison to the
	resonance peaks studied in Ref.~\cite{DaniCP22} the inter-dot tunneling is
	a factor of 2 (peak II in Ref.~\cite{DaniCP22}) to 5 (peak I) smaller, i.e.
	indicating a very weak coupling between the two dots in the work here.
	
	To analyze the spin flips, we focus on the peaks whose position and height
	changes with the magnetic field.  Figure~\ref{fig:ExpTheo}(c) shows the two
	peaks for a magnetic field $B_\perp = \SI{1.1}{T}$ and two different
	temperatures. In the further analysis the height of the peaks is analyzed
	as function of the magnetic field, i.e., the peak position on the voltage axes
	relative to $V_0$. In this way Fig.~\ref{fig:ExpTheo}(d) shows the height
	of the peaks as a function of their position for the two different
	temperatures (red circles: $T=\SI{1.5}{K}$, blue triangles: $T =
	\SI{0.1}{K}$, where peak II vanishes for higher magnetic fields). In the
	Appendix we show that also for the parallel magnetic field direction
	similar results are obtained.  Solving numerically our above model leads to
	two fitting parameters in addition to the parameters extracted for $B = 0$
	and in addition to $\Delta g$ extracted from the observed Zeeman splitting.
	The two fitting parameters are the dimensionless spin dissipation
	$\alpha_\text{spin} = 1.2\times10^{-4}$ and the gyromagnetic ratio of the
	left dot, $g_L=0.7$ (for $B_\parallel$ we obtain $g_L=1$). Only $g_L$ plays
	a role since our spin-blockade mechanism is governed by the spin relaxation
	in the left dot. The experimental results are well reproduced by our model
	(solid lines) in using these two fitting parameters for both temperatures.
	At a temperature of $T = \SI{1.5}{K}$ the model describes the experimental results
	almost perfectly showing that our assumption of an ohmic spectral density
	is well justified. At a temperature of $T = \SI{0.1}{K}$, especially at more
	negative voltages (higher magnetic fields) deviations between theory and
	experiment are observed. These deviations might be caused by the influence
	of noise on the very sharp peaks (see the Appendix for our analysis of the
	noise in the signal). Nevertheless, also at this low temperature our
	theoretical model describes the experimental results rather well.  The
	dimensionless spin dissipation $\alpha_\text{spin}$ which we obtained from
	fitting our model is our central quantity of interest, because it
	allows predictions for the spin coherence \cite{WeissPRL89,
	MakhlinRMP01}. The good description of the experimental results by our
	model clearly hints towards spin-flip cotunneling as the main process in
	our system.
	
	\section{Energy dependence of the spin decay}
	While we have seen that the spin-flip cotunneling
	predicts the behaviour in the magnetic field rather well, we now aim
	at a more direct access to the spin-flip rate as a function of the Zeeman
	spitting.
	For this purpose, we capture the scenario of the spin blockade by a rate
	equation which holds for sufficiently large magnetic field such that the
	two peaks are well separated.  We assume that an electron with arbitrary
	spin orientation enters from the source to the left dot, where it undergoes
	spin flips with the spin decay rate $\gamma$ and the corresponding thermal
	excitation rate $\bar\gamma$.  Both rates are assumed to be linked by the detailed
	balance relation $\bar\gamma(E) / \gamma(E)  = \exp(-E/kT)$.  At the peak,
	one Zeeman level is in resonance with the corresponding level of the right
	dot, the electron may proceed to the drain with some effective rate
	$\Gamma_\text{eff}$.
	The resulting rate equation provides the current
	\begin{equation}
		I(B) = \frac{2\gamma I_0}{I_0/e + 2\gamma + 2\bar\gamma}
		\label{Ianalyt}
	\end{equation}
	with $I_0$ the peak height at zero magnetic field and the Zeeman splitting
	in the left dot, $E = g_L\mu_BB$.  Interestingly, for sufficiently large
	$g_L$, this result depends on $\Omega$ and $\Gamma_R$ only via $I_0$.
	
	\begin{figure}
		\centerline{%
		\includegraphics[width=.85\columnwidth]{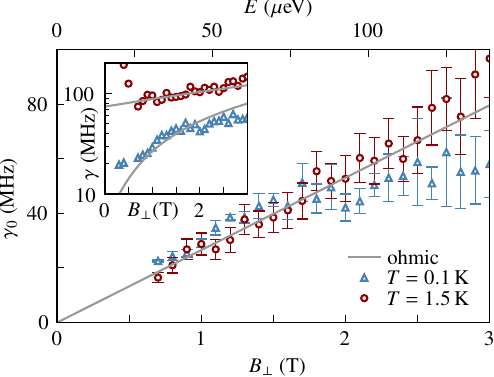}}
		\caption{Spin-flip rate in the zero-temperature limit, $\gamma_0$,
		as a function of the Zeeman splitting (upper axis) and the corresponding
		magnetic field.  The values are obtained from the approximate analytical
		solution in Eq.~\eqref{Ianalyt} together with the detailed-balance
		relation. For the size of the error bars, see the Appendix.
		Inset: Spin-flip rate at finite temperature,
		$\gamma(E) = \gamma_0(E)[1+n_\text{th}(E)]$.}
		\label{fig:effRate}
	\end{figure}
	
	Equation \eqref{Ianalyt} together with the detailed balance relation
	provides $\gamma$ as a function of the Zeeman splitting $E$ and $I_0$.  At
	low temperatures, the result corresponds to the spontaneous decay rate.
	Moreover, under the assumption that the induced decay is proportional to
	the thermal occupation of the environmental modes $n_\text{th}(E) =
	[\exp(E/kT) - 1]^{-1}$, we may compute the
	spontaneous decay rate also from measurements at higher temperatures.
	Therefore, we conjecture that $\gamma_0(E) \equiv
	\gamma(E)/[1+n_\text{th}(E)]$ is temperature independent and grows linearly
	with the Zeeman splitting and, hence, with the magnetic field. 
	Figure \eqref{fig:effRate} shows the accordingly evaluated experimental
	data.  They show a good agreement for $B_\perp\gtrsim\SI{0.5}{T}$, while below
	this value, the peaks overlap such that Eq.~\eqref{Ianalyt} does not hold.
	Especially for $T=\SI{1.5}{K}$ the agreement is quite good. One obtains spin
	relaxation rates varying between \SI{20}{MHz} at \SI{0.7}{T} and about
	\SI{80}{MHz} at \SI{3}{T}.  This linear behaviour is directly related to
	the ohmic spectral density and supports our conjecture
	of spin-flip cotunneling resolving the current blockade.
	At $T=\SI{100}{mK}$, one witnesses a deviation which we attribute to the
	already mentioned underestimation of the peak height. Moreover, for large
	Zeeman splittings, the thermal excitation rate becomes small, which
	augments the relative error.
	
	In principle, one may consider other spin-dissipation mechanisms such as
	spin--orbit or hyperfine interaction.  These, however, are expected to show
	a stronger dependence on the magnetic field and can be ruled out.
	A further conceivable mechanism to resolve the blockade is
	dissipative transitions from the left to the right dot \cite{DaniCP22}.  As
	such transitions occur also when both Zeeman levels are misaligned, they
	cannot explain the emergence of sharp resonance peaks with the observed
	asymmetry.
	
	\section{Conclusions}
	We have used a spin-dependent blockade mechanism for single electrons in
	self-assembled double quantum dots to extract spin relaxation rates
	directly from the measured resonant tunnel currents. The blockade mechanism
	here stems from inhomogeneous Zeeman splittings such that the two spin
	channels are never resonant simultaneously.  Then an electron in the
	off-resonant channel blocks transport until a spin flip occurs.  An
	analysis based on a rate equation provided the spin-flip rates which turned
	out to grow linearly with the Zeeman splitting.  Quantitatively, the
	spontaneous spin decay rate at $\SI{2}{T}$ is of the order $\SI{50}{MHz}$.
	Hence, we expect coherence times of roughly \SI{20}{ns} which corresponds to
	a few hundred coherent oscillations.
	The here discussed findings %
	identify spin-flip cotunneling as the dominating decoherence mechanism.
	
	\begin{acknowledgments}
		This work was supported by the Deutsche Forschungsgemeinschaft (DFG, German
		Research Foundation) under Germany's Excellence Strategy -- EXC 2123
		QuantumFrontiers -- 390837967, the State of Lower Saxony of Germany via the
		Hannover School for Nanotechnology, and by the Spanish Ministry of Science,
		Innovation, and Universities (Grant No.\ PID2020-117787GB-I00), and by the
		CSIC Research Platform on Quantum Technologies PTI-001.
	\end{acknowledgments}
	
	\appendix
	\asection{Data analysis}
	\subsection{Position, height, and width of the peaks}
	
	The data analysis of the experimental current-voltage characteristics consists 
	in a first step in the subtraction of the background. Therefore, the left and 
	right background to each resonance are exponentially fitted and the fits are 
	stitched together at their intersection. The combined background is then 
	subtracted from the raw current-voltage characteristics.  We refer to these 
	resulting background-less data shortly as measured data.
	
\begin{figure}[b]
		\centerline{%
		\includegraphics[]{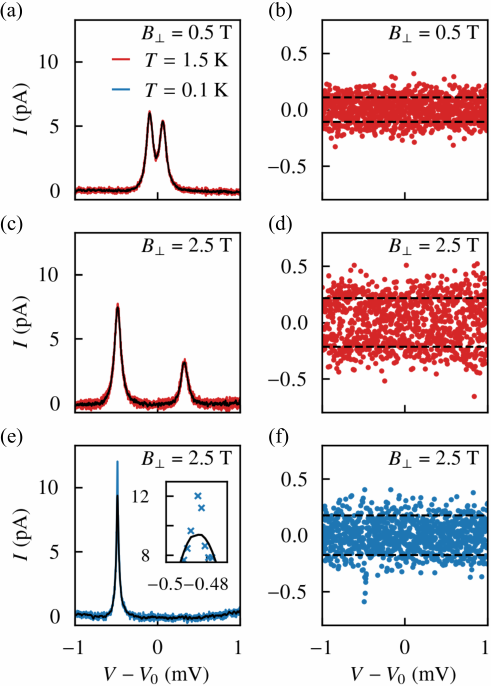}} 
		\caption{(a,c) Measured tunneling current (red line) as a function of the bias voltage
		at a temperature of $T=\SI{1.5}{K}$ for magnetic fields $B_\perp=\SI{0.5}{T}$ and
		$\SI{2.5}{T}$, respectively.  The black line is the Savitzky-Golay
		filtered signal.
		(e) Same at $T=\SI{0.1}{K}$ and $B_\perp=\SI{2.5}{T}$. The inset shows the peak
		in high resolution, while the blue lines and symbols mark measured data.
		(b,d,f) Difference of the measured data and the filtered signal.
		The dashed lines mark the range of the standard deviation.}
		\label{fig:Data_Analysis}
	\end{figure}

	To determine the peak height and position we explore the influence of the
	noise in the signal. We start by filtering the signal using a
	Savitzky-Golay filter~\cite{SavitzkyAC64} and then subtract the filtered
	signal from the measured data as shown in Fig.~\ref{fig:Data_Analysis}. The
	standard deviation of the resulting data points is converted and
	represented by the error bars in Fig.~4 in the main text. 
	At the low temperature of $T=\SI{0.1}{\kelvin}$ and higher magnetic fields,
	the peaks become very sharp [Fig.~\ref{fig:Data_Analysis}(e)] such that at
	the peak position, the deviation of the filtered signal from the measurement is
	much larger than the noise, see inset in Fig.~\ref{fig:Data_Analysis}(e).
	Therefore the sharp peak is taken into account in the size of the error
	bars. 
	
	\subsection{Fitting of the spin dissipation strength}
	
	In our data analysis, the detuning of the quantum dot (QD) levels is linked to the bias
	voltage via $\epsilon = \eta V$, where the leverage factor can be estimated
	from the assumed voltage drop across the sample as $\eta\approx0.15$.  To
	reveal the sensitivity of our fit parameters to the value of $\eta$, we
	show in the last line of Table~\ref{tab:params} how the fit parameters to
	leading order depend on $\eta$.  In particular the central quantity
	determined in this work, namely the dimensionless spin dissipation strength
	$\alpha_\text{spin}$ is practically independent of the leverage.
	
	\begin{table*}[t]
		\centering
		\caption{Model parameters determined by fitting the experimental peak
		heights for the leverage factor $\eta=0.15$.  The last row shows the
		approximate dependence of the fit parameters on $\eta$.
		Remarkably, the fit value for the dimensionless spin dissipation strength
		$\alpha_\text{spin}$ is practically independent of $\eta$ and the
		orientation of the magnetic field.
		\label{tab:params}}
		\begin{ruledtabular}
			\begin{tabularx}{\columnwidth}{ccccccc}
				& $\Gamma_R$\ \si{(\micro eV)} & $\Omega$\ (\si{\micro eV})
				& $\alpha\times 10^3$
				& $\Delta g=g_R-g_L$ & $g_L$ & $\alpha_\text{spin}\times 10^4$
				\\[.25ex]
				\colrule
				\\[-1.5ex]
				$B_\perp$         & 5 & 0.85 & 5.0 & 0.85 & 0.7 & 1.2 \\
				\\[-1.5ex]
				$B_{\parallel}$ & 5 & 0.85 & 5.0 & 1 & 1 & 1.2
				\\[.25ex]
				\colrule
				\\[-1.75ex]
				& $\propto\eta$ & $\propto\eta^{1/2}$ & $\propto\eta$ &
				$\propto\eta$ & $\propto\eta^{0}$ & $\propto\eta^0$
			\end{tabularx}
		\end{ruledtabular}
	\end{table*}
	
	\asection{Magnetic-field direction dependence}
	\begin{figure}[b]
		\centerline{%
		\includegraphics[]{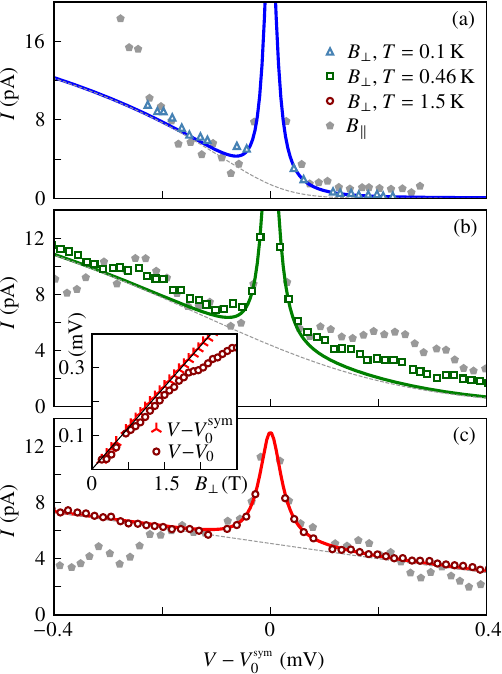}} 
		\caption{Peak heights of the measured tunneling current as a function of
		the bias voltage for perpendicular (colored symbols) and parallel (gray
		symbols) applied magnetic field at temperature
		$T=\SI{0.1}{\kelvin}$ (a), \SI{0.46}{\kelvin} (b), and
		$\SI{1.5}{\kelvin}$ (c). The peak positions are taken with respect to the
		midpoint $V_0^\text{sym}$ between the peaks. Solid and dashed
		lines, respectively, indicate the numerical data and the analytical
		approximation \eqref{Ieff} for the case of finite $B_\perp$. The inset shows the peak
		positions with respect to $V_0^\text{sym}$ and the position $V_0$ at zero
		magnetic field in dependence of $B_\perp$ for $T=\SI{1.5}{\kelvin}$. 
		}
		\label{fig:BDir}
	\end{figure}
	A magnetic field $B_\parallel$ applied parallel to the tunneling current
	(i.e.\ perpendicular to the layer)
	induces oscillations in the current-voltage characteristics which are
	absent in the case of a perpendicular field $B_\perp$ as is discussed in
	Figs.~1 and~2 of the main text. Nevertheless, the peak-to-peak distance
	increases in both cases linearly with the magnetic field. We show in
	Fig.~\ref{fig:BDir} that also the peak heights evolve overall quite
	similarly for both magnetic-field directions with the peak position when
	taking the midpoint $V_0^{\rm sym}$ between the two peaks as
	reference point. For consistency and in contrast to the main text, we here
	use also for perpendicular orientation $V_0^{\rm sym}$ as reference point.
	The inset of Fig.~\ref{fig:BDir} shows that for $B_\perp\lesssim\SI{1.7}{T}$,
	the difference between $V_0$ and $V_0^\text{sym}$ is minor.
	
	Upon closer look, one notices that for parallel magnetic field, Landau
	oscillations become apparent which for perpendicular magnetic-field
	direction is not the case. These Landau oscillations predominantly affect
	the density of states in the leads and are not captured by our theory,
	which explains the slightly larger deviation from the theoretical
	prediction in the presence of $B_\parallel$.
	
	\asection{Ohmic dissipation from spin-flip cotunneling}
	
	\begin{figure}[b]
		\centerline{%
		\includegraphics[scale=1.2]{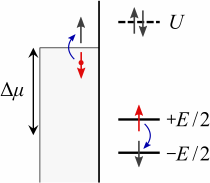}} 
		\caption{Spin-flip cotunneling in a single quantum dot coupled to a lead.
		A dot electron initially in the upper Zeeman level (red) flips to the
		spin-down level (green), while in the leads a spin-down electron at the
		Fermi surface is excited under spin flip. The virtually populated dot states of this
		second-order process may be the empty dot or the singlet state which is
		separated from the singly occupied state by the interaction energy $U$.}
		\label{fig:cotunnel}
	\end{figure}
	
	The coupling of a spin to a bosonic heat bath can be written as $V =
	\hbar\sigma_x\xi$ with the effective bath coordinate $\xi =
	\sum_\nu\lambda_\nu(a_\nu + a^\dagger_\nu)$, where $a_\nu$ is the usual
	bosonic annihilation operator of a bath mode with frequency $\omega_\nu$
	and the coupling energy $\hbar\lambda_\nu$.  Its auto-correlation function
	$C(t) = \langle\xi(0)\xi(t)\rangle$ contains all information about the
	bath needed to evaluate the dissipative terms of the Bloch-Redfield
	equation \cite{RedfieldIBMJRD57, Blum1996}. For a Gibbs state, in
	frequency space it reads
	\begin{equation}
		C(\omega) = 2 J(\omega) n_\text{th}(\hbar\omega)
		\label{app:Cohmic}
	\end{equation}
	with the bath spectral density $J(\omega) = \pi\sum_\nu |\lambda_\nu|^2
	\delta(\omega-\omega_\nu)$ and the bosonic thermal occupation number
	$n_\text{th}(E) = [\exp(E/kT)-1]^{-1}$.  In the following, we demonstrate
	that spin-flip cotunneling corresponds to quantum noise obeying the
	correlation \eqref{app:Cohmic} with ohmic spectral density, i.e.,
	$J(\omega)\propto\omega$.
	
	To this end, we employ ideas developed in the context of Pauli spin
	blockade in double quantum dots \cite{QassemiPRL09, CoishPRB11}.  Focussing
	on spin flips within a QD, we consider a single orbital with Zeeman
	splitting $E$ coupled to a lead as is sketched in Fig.~\ref{fig:cotunnel}.
	We assume that the dot orbital lies significantly below the Fermi surface of
	the lead such that $E \ll \Delta\mu, kT$. Moreover, the intra-dot
	interaction $U$ allows only virtual double occupation of the QD.
	
	A convenient starting point is the $T$-matrix formulation of the dot-lead
	tunneling process for the tunneling Hamiltonian $H_T = \sum_{k,\sigma}
	t_k(c^\dagger_{k\sigma} d_\sigma + d^\dagger_\sigma c_{k\sigma})$ for
	spin-conserving tunneling between the QD and the lead with the
	spin-independent matrix elements $t_k$.  The corresponding electron
	annihilation operators are $c_{k\sigma}$ and $d_\sigma$, where $\sigma$ is
	the spin quantum number and $k$ labels the lead orbitals.  In the situation
	considered, the Pauli exclusion principle together with Coulomb repulsion
	does not allow resonant tunneling. Then the leading contribution of the
	$T$-matrix is the second-order term \cite{BruusFlensberg2004, CoishPRB11}
	\begin{equation}
		T = H_T \frac{1}{E_i-H_0} H_T,
	\end{equation}
	which replaces the tunnel Hamiltonian in the Bloch-Redfield equation and in
	the golden-rule rates.  Here $E_i$ is the energy of an initial state
	$|i\rangle$ in the many-electron Hilbert space of QD and lead, while $H_0$
	is the Hamiltonian in the absence of tunneling.  It has to be
	evaluated for the virtually occupied intermediate states $|v\rangle$
	created by $H_T$.
	
	Following Ref.~\cite{CoishPRB11}, we estimate $E_i-H_0$
	for the initial QD state $|\up\rangle$.  Its decay towards
	$|\down\rangle$ induced by $H_T$ may occur via the intermediate states
	$c^\dagger_{k\up}d_\up|i\rangle$ and $d_\down^\dagger c_{k\down}|i\rangle$,
	i.e., the empty and the doubly occupied QD.  Since the total energy of
	the transition induced by $T$ must be conserved, the necessary
	excitation of a lead electron can occur only at the Fermi surface.
	Therefore the energy differences of the initial state and the intermediate
	states are $E_i-E_v=-\Delta\mu$ and $E_i-E_v=-U$, respectively, see
	Fig.~\ref{fig:cotunnel}. Hence the $T$-matrix can be approximated as
	\begin{equation}
		T = \hbar ( d^\dagger_\uparrow d_\downarrow \zeta
		+ d^\dagger_\downarrow d_\uparrow \zeta^\dagger )
		\label{app:Teff}
	\end{equation}
	with the (operator-valued) quantum noise
	\begin{equation}
		\zeta = \frac{1}{\hbar W} \sum_{k,k'} t_k t_{k'} c^\dagger_{k\downarrow}
		c_{k'\uparrow}
		\label{app:eta}
	\end{equation}
	and the effective energy difference $W$ defined via
	\begin{equation}
		\frac{1}{W} = \frac{1}{\Delta\mu} + \frac{1}{U} .
		\label{app:W}
	\end{equation}
	A straightforward calculation yields for a Gibbs state of the lead with
	spin-independent chemical potential $\mu$ the correlation function
	\begin{align}
C_\text{cot}(t)
={} & \langle\zeta^\dagger(0)\zeta(t)\rangle = \langle\zeta(0)\zeta^\dagger(t)\rangle
\\
={} & \Big(\frac{\Gamma}{2\pi\hbar W}\Big)^2 \int d\epsilon\,d\epsilon'
 e^{-i(\epsilon-\epsilon')t/\hbar}
\\ & \hspace{11ex}\times 
f(\epsilon-\mu) [1-f(\epsilon'-\mu)]
\nonumber
	\end{align}
	with the Fermi function $f(E) = [\exp(E/kT)+1]^{-1}$.  In a continuum
	approximation, we have defined $\Gamma = 2\pi\sum_k |t_k|^2
	\delta(\epsilon-\epsilon_k)$, where here $\Gamma/\hbar$ is the dot-lead tunnel
	rate at the Fermi surface.  By substituting
	$\epsilon\to\hbar\omega+\epsilon'$ and evaluating the $\epsilon'$-integral,
	we obtain the auto-correlation in frequency space reading
	\begin{equation}
		C_\text{spin}(\omega) = \pi\alpha_\text{spin} \omega\, n_\text{th}(\hbar\omega) ,
		\label{app:Cw}
	\end{equation}
	with the dimensionless spin-dissipation strength
	\begin{equation}
		\alpha_\text{spin} = \Big(\frac{\Gamma}{\pi W}\Big)^2 .
		\label{app:alpha-spin}
	\end{equation}
	Thus the quantum noise entailed by cotunneling on a spin has the
	second-order correlation of a bosonic heat bath with ohmic spectral density
	$J_\text{spin}(\omega) = \pi\alpha_\text{spin}\omega/2$.
	
	The coupling operator \eqref{app:Teff} between the spin and the noise
	together with the correlation function \eqref{app:Cw} allows us to evaluate
	the dissipation kernel of the Bloch-Redfield equation, such that spin-flip
	cotunneling can be treated numerically.  Moreover, it provides the spin
	decay rate \cite{WeissPRL89, MakhlinRMP01}
	\begin{equation}
		\gamma(E) = \frac{\pi}{\hbar} \alpha_\text{spin}E\, [n_\text{th}(E)+1]
		\label{app:gamma}
	\end{equation}
	with the Zeeman splitting $E = g\mu_B B$ and the Bohr magneton $\mu_B$.
	The corresponding thermal excitation rate $\bar\gamma(E) = \gamma(E)
	\exp(-E/kT)$ obeys the detailed-balance relation between emission and
	absorption.
	
	As discussed in the main text, our experimental results are compatible with
	a dissipation strength of the order $\alpha_\text{spin}\sim 10^{-4}$, which
	results from Eqs.~\eqref{app:W} and \eqref{app:alpha-spin} for the quite realistic values
	$\Gamma = \SI{100}{\micro eV}$ and $U\approx\Delta\mu \approx
	\SI{1.5}{meV}$.  In experiments with (almost) closed QDs, $\Gamma$
	and hence $\alpha_\text{spin}$ may be significantly smaller, such that spin
	decoherence is governed by other mechanisms such as hyperfine interaction.
	
	Since $\zeta\neq\zeta^\dagger$, the $T$-matrix in Eq.~\eqref{app:Teff} is
	not exactly of the form as in the spin-boson model, which is
	$\propto\sigma_x\xi$ with the hermitian operator-valued noise $\xi$.
	Therefore, we introduce the hermitian and statistically independent noise
	operators $\xi_x = (\zeta^\dagger+\zeta)/2$ and $\xi_y =
	(\zeta^\dagger-\zeta)/2i$.  Their auto-correlation function is also of the
	form \eqref{app:Cohmic} with $J(\omega) = \pi\alpha_\text{spin}\omega/4$.
	Then  the $T$-matrix (concisely written with Pauli matrices
	$\sigma_+ = d^\dagger_\up d_\down = \sigma_-^\dagger$) becomes
	\begin{equation}
		T = \sigma_-\zeta^\dagger + \sigma_+\zeta
		= \sigma_x\xi_x + \sigma_y\xi_y .
	\end{equation}
	This implies that on the level of Bloch-Redfield theory, spin-flip
	cotunneling causes the same spin dissipation as two equal bosonic
	heat baths coupled with strength $\alpha_\text{spin}/2$ via
	$\sigma_x$ and $\sigma_y$, respectively.  We have verified that
	coupling to either bath with double strength yields the practically same
	results for the current.
	
	\asection{Analytical approximation}
	\begin{figure}
		\centerline{%
		\includegraphics[scale=1.2]{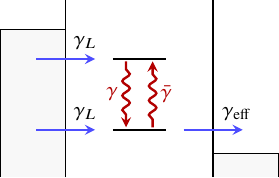}} 
		\caption{Effective model underlying the rate equation \eqref{MEeff}.  The
		right QD and the drain are replaced by an effective depletion rate
		$\gamma_\text{eff}$ for the lower Zeeman level, while the upper Zeeman
		level is not directly connected to the drain.  The dissipative spin
		dynamics is captured by incoherent spin flips, while coherences are
		neglected.}
		\label{fig:effModel}
	\end{figure}
	
	To obtain an analytical approximation for current, we derive a rate
	equation in which the tunneling from the left QD to the right QD and from
	there to the drain is replaced by an effective rate. Together with the
	spin-flip rates $\gamma$ and $\bar\gamma$, this provides the result
	visualized in Fig.~\ref{fig:BDir} by dashed lines.
	
	When the Zeeman levels of the spin-down electrons on both QDs are aligned,
	an electron with this spin projection entering from the source can
	resonantly tunnel to the drain with some effective rate
	$\gamma_\text{eff}$.  By contrast, a spin-up electron will get stuck in the
	left QD.  Owing to Coulomb repulsion, it blocks the resonant spin-down
	channel. This blockade can be resolved by a spin flip which we assume to
	happen with a rate $\gamma$.  Once the electron is in the spin-down
	channel, we are back to the former situation.  For consistency, we also
	have to take into account thermal spin excitation from the lower Zeeman
	level to the upper one, which occurs with rate $\bar\gamma$.
	
	This scenario is sketched in Fig.~\ref{fig:effModel} and can be described
	by the master equation
	\begin{equation}
		\frac{d}{dt}
		\begin{pmatrix}		p_0\\p_\uparrow\\p_\downarrow \end{pmatrix}
		=
		\begin{pmatrix}			-2\gamma_L & 0 & \gamma_\text{eff} \\
			\gamma_L & -\gamma & \bar\gamma \\
		\gamma_L & \gamma & -\bar\gamma -\gamma_\text{eff} \end{pmatrix}
		\begin{pmatrix}		p_0\\p_\uparrow\\p_\downarrow \end{pmatrix}		,
		\label{MEeff}
	\end{equation}
	where $\gamma_L = \Gamma_L/\hbar$.
	The current is given by the population of the lower Zeeman level times the
	effective decay rate to the drain,
	\begin{equation}
		I = e\gamma_\text{eff} \, p_\downarrow.
	\end{equation}
	In the absence of the spin-up channel,
	the physical situation is that of spin independent transport. Therefore, we
	can identify the corresponding current with the one in the absence of the
	magnetic field, $I_0 = e\gamma_\text{eff}$.
	It is straightforward to obtain $p_{\downarrow}$ from the master equation
	\eqref{MEeff}, such that in the limit $\gamma_L \gg
	\gamma,\bar\gamma,\gamma_\text{eff}$, the current reads
	\begin{equation}
		I = \frac{2\gamma I_0}{I_0/e + 2\gamma +2\bar\gamma}.
		\label{Ieff}
	\end{equation}
	Since we have assumed clearly distinct Zeeman levels, this result holds
	only when the peaks for the two spin channels do not overlap, i.e., far
	from the central peak at $B=0$.  Interestingly, this expression does not
	depend on the effective tunnel rate to the drain, $\gamma_\text{eff}$.

\end{document}